\documentclass[a4paper,12pt]{article}
\usepackage[affil-it]{authblk}

\usepackage{amsfonts,amsthm,amsmath,amssymb,graphicx,hyperref,youngtab}
\usepackage{empheq}
\usepackage{cite}
\usepackage{multirow}
\usepackage{booktabs}
\usepackage{rotating}
\usepackage{graphics}
\usepackage[usenames,dvipsnames,svgnames,table]{xcolor}

\makeatletter
\newcommand{\vast}{\bBigg@{1}}
\newcommand{\Vast}{\bBigg@{5}}
\makeatother

\usepackage{hyperref}
\hypersetup{
colorlinks,breaklinks,
            urlcolor=[rgb]{0.8,0.2,0.5},
            citecolor=[rgb]{0.6,0,0},
}

\numberwithin{equation}{section}

\newcommand{\be}{\begin{equation}}
\newcommand{\ee}{\end{equation}}
\newcommand{\bea}{\begin{eqnarray}}
\newcommand{\eea}{\end{eqnarray}}

\newcommand{\nn}{\nonumber}
\newcommand{\Tr}{\text{Tr}}

\newcommand{\pd}{\partial}

\begin{document}

\title{Entanglement and mutual information in two-dimensional nonrelativistic field theories}

\author{Seyed Morteza Hosseini%
  \thanks{Electronic address: \texttt{morteza.hosseini@mib.infn.it}}}
\affil{Dipartimento di Fisica, Universit\`a di Milano-Bicocca, I-20126 Milano, Italy}
\affil{INFN, sezione di Milano-Bicocca, I-20126 Milano, Italy}

\author{\'{A}lvaro V\'{e}liz-Osorio%
  \thanks{Electronic address: \texttt{alvaro.velizosorio@wits.ac.za}}}
\affil{Mandelstam Institute for Theoretical Physics, School of Physics\\University of the Witwatersrand, WITS 2050, Johannesburg, South Africa}

\date{Dated: \today}

\maketitle

\begin{abstract}
We carry out a systematic study of entanglement entropy in nonrelativistic conformal field theories via holographic techniques. After a discussion of recent results concerning Galilean conformal field theories, we deduce a novel expression for the entanglement entropy of (1+1)-dimensional Lifshitz field theories --- this is done both at zero and finite temperature. Based on these results, we pose a conjecture for the anomaly coefficient of a Lifshitz field theory dual to new massive gravity.
It is found that the Lifshitz entanglement entropy at finite temperature displays a striking similarity with that corresponding to a flat space cosmology in three dimensions. We claim that this structure is an inherent feature of the entanglement entropy for nonrelativistic conformal field theories. We finish by exploring the behavior of the mutual information for such theories.
\end{abstract}

\hypersetup{
colorlinks,breaklinks,
            linkcolor=black
}
\newpage
\tableofcontents

\hypersetup{
colorlinks,breaklinks,
            linkcolor=[rgb]{0,0,0.7}
}
\section{Introduction}

The first concrete realization of the fascinating {\it holographic principle}
appeared in the context of string theory \cite{Maldacena:1997re}
and has led to a number of interesting developments.
Maldacena\rq{}s conjecture, or the AdS/CFT correspondence,
in its original form relates the strongly interacting
${\cal N}=4$ super Yang-Mills theory with gauge group $U(N)$ to
type IIB string theory on an $\text{AdS}_5\times {\mathbb S}^5$ background.
Both sides of the duality are endowed with a superconformal $SU(2,2|4)$ symmetry group.
This matching of symmetry is the first step towards establishing the correspondence.
The AdS/CFT correspondence relates systems that are different in many aspects
by providing a precise dictionary between quantities and phenomena in one system to those in
the other.  More precisely, with the help of the correspondence,
we can formulate questions regarding quantum gravity in asymptotically AdS (AAdS)
spacetimes as problems in a lower-dimensional gauge theory, and vice versa.

\medskip
One of the interesting offshoots of the aforementioned duality
is the application of holographic models to study condensed matter systems\cite{Taylor:2008tg,Balasubramanian:2008dm,Kachru:2008yh,Hartong:2014oma,Korovin:2013bua,Griffin:2012qx,Son:2008ye}. In condensed matter applications, one is often interested in theories that
display anisotropies between space and time dimensions; i.e., non-Lorentz invariant.
For instance, in many condensed matter systems, one finds phase transitions governed
by fixed points which exhibit dynamical scaling instead of the more familiar
scale invariance which arises in the conformal group.
Inspired by the symmetry matching discussed above,
one searches for dual background geometries which capture 
the symmetries of strongly coupled nonrelativistic conformal field theories. 
Thus, one is led to consider spacetimes that are asymptotically Lifshitz (ALif) or Schr\"odinger.

\medskip
It is the goal of this work to perform a detailed study of certain properties of
(1+1)-dimensional nonrelativistic conformal field theories (CFT).
Our main focus is on theories with Lifshitz and Galilean conformal symmetries.
Below, we shall study the entanglement entropy (EE) and
mutual information for simple subsystems in such theories.
Roughly speaking, the EE quantifies our ignorance
once access to the full system is lost. Meanwhile,
the mutual information associated to two subsystems provides an upper bound
on the correlations between operators each supported on one of the subsystems. 
Analytic computations of EE and mutual information can be rather hard in general settings.
Nonetheless, there is a great body of literature with many
interesting results for a number of quantum field theories (QFT),
see, e.g., \cite{Casini:2009sr,Calabrese:2004eu}.
Lately, there has been a renewed interest in the study of EE from people working on holography and gravity. 
This is thanks to a reformulation of the problem, in the light of the AdS/CFT correspondence,
which allows one to compute the EE in a succinct manner \cite{Ryu:2006bv}.
It is in this spirit that we approach the above questions.
We hope that the present explorations might provide insights into
how quantum information is encoded in nonrelativistic CFT's.

\medskip
Three-dimensional gravity has played a leading role in addressing a number of important conceptual issues in quantum gravity. In many respects,
the AdS$_3$/CFT$_2$ correspondence is the best understood example of the gauge/gravity duality.
One of the benefits of working in three dimensions is that we have a great deal of analytic control and exact statements can be made. 
In particular, results in holographic EE can be verified explicitly by matching them to QFT results. 
Three-dimensional theories of gravity with higher-derivative interactions have been the subject of many investigations  in recent years. 
It is a well known fact that three-dimensional gravity has no local degrees of freedom. By endowing the graviton with a mass it can be made to carry two propagating degrees of freedom.
In this work we study one such theory, the so-called new massive gravity (NMG) \cite{Bergshoeff:2009hq}. There are a number of interesting spacetimes supported by this theory with no counterpart in pure Einstein gravity \cite{Fareghbal:2014kfa, Oliva:2009ip}. In particular, NMG admits ALif solutions amongst which black holes can be found \cite{AyonBeato:2009nh} --- these solutions will play a central role in this article. 

\medskip
In this work, we shall present a novel expression for the EE
corresponding to a class of Lifshitz field theories in 1+1 dimensions (Lif$_2$),
both at zero and finite temperature. As far as we are aware, 
this is the first time that such formulas have been constructed.
The expressions in question, display the appropriate anisotropic scaling behavior with respect to the
system's temperature. Moreover, the finite temperature Lifshitz EE displays an eye-catching structural similarity with the EE 
for field theories dual to flat space cosmologies (FSC) \cite{Hosseini:2015uba}. The deviations of the Lif$_2$ and FSC EE's with respect to the CFT EE at finite temperature, might be a consequence of the ultralocality exhibited by certain nonrelativistic fixed points \cite{Freedman:2005,Yang:2004,Sachdev:1996,Ghaemi:2004}.
Comparison of our general formula for the Lif$_2$ EE with those corresponding to CFT$_2$ and
Galilean conformal filed theories in 1+1 dimensions (GCFT$_2$) leads us
to conjecture a formula for the anomaly coefficient of a Lif$_2$ field theory.
Finally, with these similarities in mind, we discuss the general form 
of the phase diagram for the mutual information in nonrelativistic CFTs.  We hope that
the outcomes of this paper may provide a valuable bridge between gravity and condensed matter physics.

\medskip
The layout of this paper is as follows:
Section \ref{EE and holography} contains details of the holographic approach
for calculating the EE corresponding to field theories dual to Einstein gravity.
Then we proceed to the computation of holographic the EE in higher-derivative gravity theories.
We also clarify our notations.
In section \ref{EE for FSC} we revisit the FSC geometry and its holographic EE.
We start section \ref{EE for ALif} by studying Lifshitz vacua and black holes in NMG.
We perform an exact analysis of the holographic EE both at zero and finite temperature corresponding to these spacetimes.
We go on to comment on the putative dual field theory to ALif geometries
and propose a conjecture for the Lif$_2$ anomaly coefficient.
With all the ingredients in hand, we move to compute the
mutual information for a class of nonrelativistic field theories and interpret our finding in section \ref{MI}.
We conclude in section \ref{Conclusions and outlook} with a summary of our results and discuss possible future directions.
In Appendix \ref{AppA} we display massive gravity action in 2+1 dimensions.
Appendix \ref{AppB} and \ref{AppC} are devoted to nonrelativistic conformal algebra.

\paragraph*{Note added:} After this paper is published we noticed a mistake in our Mathematica code.
The geodesic in the case of Lifshitz spacetimes does not solve the differential equations coming from the variation of \eqref{EE H1}.
This is against the claim in the main text. This has been also pointed out in the recent paper \cite{Basanisi:2016hsh}.

\section{Entanglement entropy and holography}
\label{EE and holography}

In this section, we provide a brief review of the basic techniques employed to compute EE holographically. 
We simply summarize the basic formulas and refer the reader to the literature for more details. We start with a discussion of the procedure that must be followed when dealing with field theories dual to Einstein gravity -- namely, the Ryu-Takayanagi prescription. Then, we present the strategy and formulas needed to perform this calculation for higher curvature theories of gravity.

\subsection{The Ryu-Takayanagi prescription}

\begin{figure}
    \centering
    \includegraphics[scale=0.45]{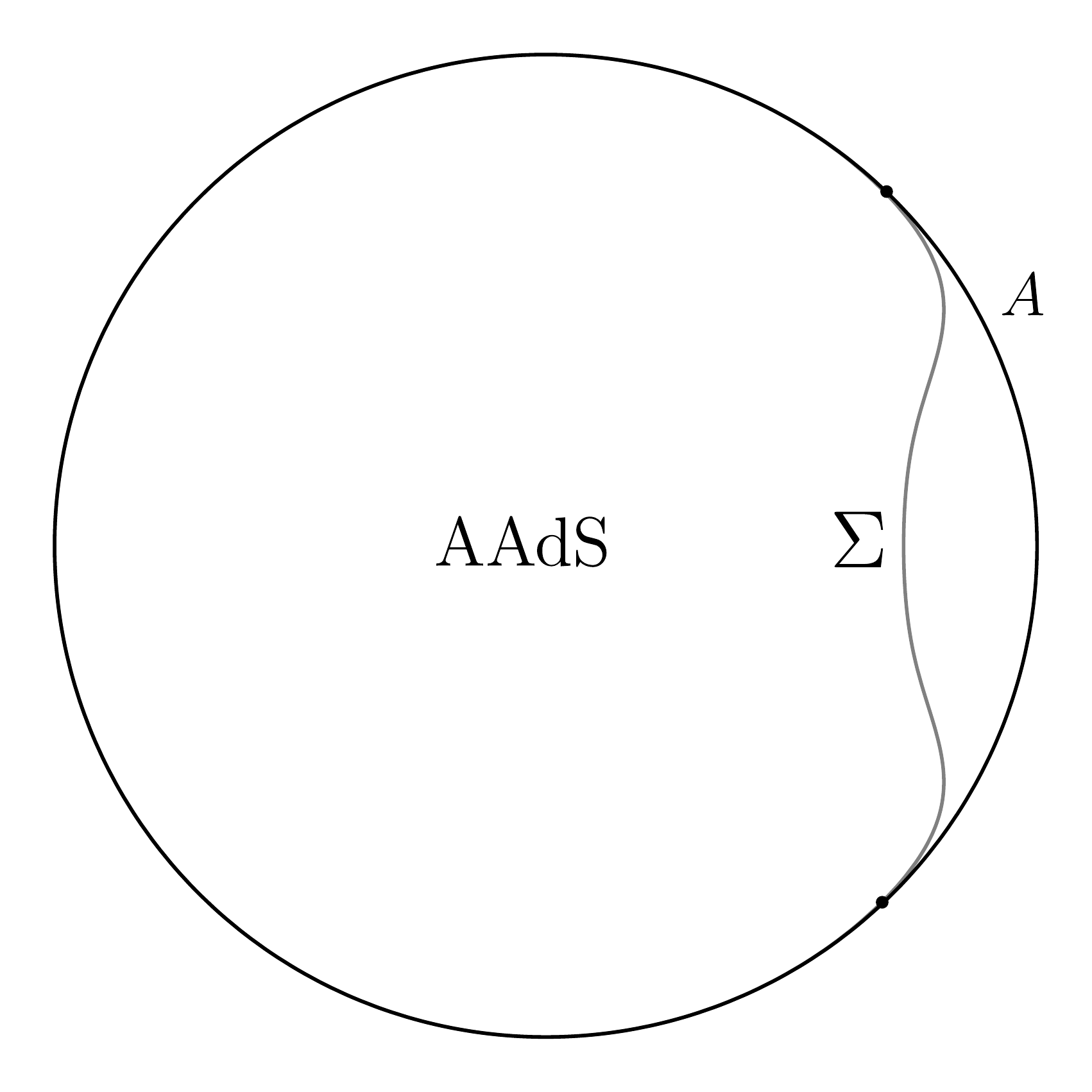}
\caption{The Ryu-Takayanagi prescription; $A$ is the entangling region, and $\Sigma$ is the extremal surface in the asymptotically AdS background, where we have used the map $r \to \arctan (r)$
to compactify the radial coordinate.}\label{RTf}
\end{figure}

In recent years, the study of EE has attracted a great deal of interest from the hep-th community,
this is partly thanks to the holographic reformulation of the problem due to Ryu and Takayanagi (RT) \cite{Ryu:2006bv}.
The RT prescription recasts he computation of EE
as a Plateau problem in an AAdS spacetime.
More concretely, in order to calculate the EE for a subsystem $A$ in the boundary theory, one needs to extremize the functional
\be\label{RT1}
S(A)=\frac{1}{4G}
\int_{\Sigma} d^{d-1}y\sqrt{h}\, ,
\ee
where $\Sigma$ is a co-dimension two hypersurface anchored
at $\partial A$ (see Fig.\;\ref{RTf}), and $h$ is the induced metric on $\Sigma$. To find  $h$, we proceed as follows,
first, we find a basis for the vector space normal to the surface $\Sigma$.
In particular,  we choose basis vectors $n_{(0)}^\mu$ and $n_{(1)}^\mu$ such that 
\be 
g_{\mu\nu}n_{(\alpha)}^\mu n_{(\beta)}^\nu=\eta_{\alpha\beta}\, ,
\ee
where $\eta_{\alpha\beta}$ is the two-dimensional Minkowski metric.
In terms of these vectors, 
the induced metric is simply
\be \label{ind}
h_{\mu\nu}= g_{\mu\nu}-\eta^{\alpha\beta}\left(n_{(\alpha)}\right)_\mu\left(n_{(\beta)}\right)_\nu\, .
\ee
\medskip
Once we substitute \eqref{ind} into the functional \eqref{RT1} and perform the variation, we must solve an ordinary differential 
equation (ODE)\footnote{For static spacetimes.}. Solving it, and plugging its solution back into the functional \eqref{RT1} we find the holographic EE 
\be\label{RT2}
S_{{\rm EE}}(A)=\frac{1}{4G}\min_{\partial A=\partial \Sigma} \text{Area}\left(\Sigma\right).
\ee
Indeed, as shown in \cite{Ryu:2006bv} when $g_{\mu\nu}$ is the radius $L$ Poincar\'e-$ \rm{AdS}_3$ metric
\be\label{PAdS}
ds^2=-\frac{r^{2}}{L^{2}}dt^2
+\frac{L^2}{r^2} dr^2
+\frac{r^2}{L^2} dx^2\, ,
\ee
and $A$ is an interval of length $\ell$,
Eq.\,\eqref{RT2} yields
\be \label{CC1}
S_{{\rm EE}}(A)=\frac{c}{3}\log\left(\frac{\ell}{\delta}\right)\, ,
\ee
where $\delta$ is a UV cutoff.
This result matches precisely $\rm{CFT}_2$ computations \cite{Calabrese:2004eu, Holzhey:1994we}. Furthermore, if the bulk is corresponds to a BTZ black hole 
\be\label{BTZ}
ds^2=-\frac{r^{2}}{L^{2}}\left(1-\frac{8 M G L^2}{r^2}\right)dt^2
+\frac{L^2}{r^2}\left(1-\frac{8 M G L^2}{r^2}\right)^{-1}dr^2\,
+r^2 d\phi^2\, ,
\ee
the RT prescription gives
\be \label{CC2}
S_{{\rm EE}}(A)=\frac{c}{3}\log\left[\frac{\beta}{\pi\delta}\sinh\left(\frac{\pi \ell}{\beta}\right)\right],
\ee
where $\beta$ is the inverse Hawking temperature of the black hole; this outcome is also in perfect agreement with 
$\rm{CFT}_2$ expectations \cite{Calabrese:2004eu}. 

\subsection{Higher curvature entanglement}

Notwithstanding its great efficacy, the prescription (\ref{RT2}) has an important limitation. 
Experience with black hole entropy has taught us that in
the presence of higher curvature corrections, black hole entropy computations must be modified \cite{Wald:1993nt} . 
Since EE and black hole entropy are deeply related,
it is clear that this insight should also apply to the RT prescription.
This question has recently been addressed by many authors 
\cite{Bhattacharyya:2013gra, Camps:2013zua, Dong:2013qoa,Fursaev:2013fta, Solodukhin:2008dh}.
In the following we summarize results that will be useful in the forthcoming sections.

\medskip
For concreteness, let us consider a general four-derivative gravity action
\be\label{four der}
S=\frac{1}{16\pi G}\int d^{d+1}x \sqrt{-g}\left[R-2\lambda+c_1 R^2+c_2 R_{\mu\nu}R^{\mu\nu}+c_3 R_{\mu\nu\rho\sigma}R^{\mu\nu\rho\sigma}\right].
\ee
The EE for its field theory duals is still determined via a co-dimension
two surface $\Sigma$ extending into the bulk as in Fig.\,\ref{RTf}.
However, the shape that $\Sigma$ ought to take in order to yield the correct
value is now determined by extremizing the functional \cite{Fursaev:2013fta, Camps:2013zua,Dong:2013qoa}
\be\label{EE HC}
S_{{\rm EE}}=\frac{1}{4G}
\int_{\Sigma} d^{d-1}y\sqrt{h}
\left[1+2 c_1 R +c_2\left(R_{||}-\frac{1}{2}{\cal K}^{2}\right)+2 c_3\left(R_{||\,||}-\Tr\left({\cal K}\right)^{2}\right)
\right]\, ,
\ee
instead of (\ref{RT1}).
Below, we explain how to obtain the different geometric quantities entering into this functional.
The contributions from the ambient Riemann curvature in Eq.\,\eqref{EE HC} read
\be
 R_{||\,||}=\eta^{\alpha\delta}\eta^{\beta\gamma}\left(n_{(\alpha)}\right)^\mu\left(n_{(\delta)}\right)^\nu \left(n_{(\beta)}\right)^\rho\left(n_{(\gamma)}\right)^\sigma R_{\mu\nu\rho\sigma}\, ,
\ee
and
\be
 R_{||}=\eta^{\alpha\beta}\left(n_{(\alpha)}\right)^\mu\left(n_{(\beta)}\right)^\nu R_{\mu\nu}\, .
\ee
While the extrinsic curvature is given by
\be
\left({\cal K}_{(\alpha)}\right)_{\mu\nu}
=h_{\mu}^{\;\lambda}h_{\nu}^{\;\rho}\nabla_\rho \left(n_{(\alpha)}\right)_\lambda\, , 
\ee
where $\nabla$ is the covariant derivative with respect to the bulk metric.
The contractions of ${\cal K}_{(\alpha)}$ entering the functional  (\ref{EE HC}) can be written as
\be\label{K sq}
{\cal K}^2\equiv \eta^{\alpha\beta}( {\cal K}_{(\alpha)})_{\mu}^{\;\mu}( {\cal K}_{(\beta)})_{\nu}^{\;\nu}\, ,
\ee
and
\be
\Tr\left({\cal K}\right)^2\equiv \eta^{\alpha\beta}( {\cal K}_{(\alpha)})_{\mu}^{\;\nu}( {\cal K}_{(\beta)})_{\nu}^{\;\mu}\, .
\ee

\medskip
For instance, if the underlying theory were NMG (\ref{massiveAc}), then
in the notation of Eq. (\ref{four der}) we would have 
\be 
c_1=-\frac{3}{8m^2}\, ,\qquad c_2=\frac{1}{m^2}\, ,\qquad c_3=0\, .
\ee
Hence, the functional (\ref{EE HC}) reduces to
\be\label{EE H1}
S_{{\rm EE}}=\frac{1}{4G}
\int_\Sigma d\tau\sqrt{h}
\left[1+\frac{1}{m^2}\left(R_{||}-\frac{1}{2}{\cal K}^{2}-\frac{3}{4}R\right)
\right]\, .
\ee
Notice that the variational problem associated with this functional yields fourth-order ODE's.
Thus, we must be able to provide four boundary conditions to find the EE via Eq.\;\eqref{EE H1},
this problem was addressed in \cite{Hosseini:2015vya} leading to the proposal of the \emph{free-kick conditions}.
Employing these conditions and the functional \eqref{EE H1},
it can be shown that the EE for the Oliva-Tempo-Troncoso (OTT) black hole \cite{Oliva:2009ip}
takes the correct form \eqref{CC2}, see \cite{Hosseini:2015vya} for more details.

\section{Entanglement entropy for flat space cosmology}
\label{EE for FSC}
Before embarking in the computation of the EE for asymptotically Lifshitz (ALif) spacetimes,
let us review some recent results concerning the EE of GCFT\rq{}s in 1+1 dimensions. 
These are theories that posses an infinite dimensional symmetry algebra, the Galilean conformal algebra (GCA),
Eq.\,\eqref{GCA}. Interestingly, the GCA is isomorphic to the
symmetry algebra of asymptotically flat (AFlat) spacetimes at null infinity.
This algebra, known as the Bondi-Metzner-Sachs (BMS) algebra
\cite{Bondi:1962px, Sachs:1962wk, Sachs:1962zza,Barnich:2010eb},
is generated by the direct sum of the infinitesimal diffeomorphisms on the circle (${\cal J}_m$)
with an Abelian ideal of supertranslations (${\cal P}_m$). This isomorphism has led to the proposal of a 
duality between gravity in AFlat spacetimes and GCFT's in one dimension less, known as 
the BMS/GCA correspondence \cite{Bagchi:2010zz,Bagchi:2009my, Bagchi:2012cy}.

\medskip
To get a handle on the EE for a class of $\rm{GCFT}_2$'s, we consider
solutions to topological massive gravity (TMG) \cite{Deser:1982vy},
whose action is given by Eq.\,\eqref{massiveAc} in the limit when $m\to\infty$.
Such a theory necessarily breaks parity.
The $\rm{BMS}_3$ algebra corresponding to solutions of TMG reads \cite{Barnich:2006av, Bagchi:2012yk},
\begin{align}\label{BMS}
   &[{\cal J}_m,{\cal J}_n]=(m-n){\cal J}_{n+m}+\frac{1}{4G\mu}m(m^2-1)\delta_{m+n,0}\, , \nn\\
   &[{\cal J}_m,{\cal P}_n]=(m-n){\cal P}_{n+m}+\frac{1}{4G}m(m^2-1)\delta_{m+n,0}\, ,\\
  & [{\cal P}_m,{\cal P}_n]=0\, ,\nn
\end{align}
which evidently corresponds to a $\rm{GCA}_2$ with central charges
\be
c_{LL}=\frac{3}{G \mu }\, ,\qquad
c_{LM}=\frac{3}{G}\, .\label{TMG cc}
\ee
Notice that in the Einstein gravity limit ($\mu\rightarrow\infty$) $c_{LL}=0$.
Having a GCA with two non-vanishing central extensions is one of the attractive features of TMG. The quest for holographic duals of field theories with independent central charges is an intersting direction to explore.

\medskip
At first sight, it might appear that there are no interesting purely gravitational configurations 
to study flat-space holography in 2+1 dimensions. Indeed, in order to obtain AFlat black holes 
in 2+1 dimensions one needs to add matter in such a way that the dominant energy condition is violated. 
Nevertheless, there is an interesting limit of rotating Ba\~nados-Teitelboim-Zanelli (BTZ)
black holes in which one can obtain FSCs \cite{Cornalba:2002fi}.
These are time-dependent solutions of (2+1)-dimensional gravity theories,
such as Einstein gravity and topologically massive gravity (TMG),
describing expanding (contracting) universes with flat asymptotics. 
Moreover, they are endowed with a non-trivial Bekenstein-Hawking entropy associated to their cosmological horizons.

\medskip
Alternatively, the FSC can be viewed as a shifted-boost orbifold of Minkowski spacetime, and its line element reads
\begin{align}
ds_{\rm FSC}^2&=\hat r_{+}^{2}\left(1-\frac{r_{0}^2}{r^2}\right)d\tau^{2}
-\frac{dr^2}{\hat r_{+}^{2}\left(1-\frac{r_{0}^2}{r^2}\right)}
+r^2\left(d\phi-\frac{\hat r_{+}r_{0}}{r^2}d\tau\right)^2\, ,
\end{align}
where 
\be
\hat r_{+}=\sqrt{8GM}\, ,\qquad r_{0}=\sqrt{\frac{2G}{M}}|j|\, ,\label{para}
\ee
where $j$ is a constant related to the angular momentum by
\be
 J=j-\frac{M}{\mu}\, ,
\ee
and $M$ is the mass of the solution.
Upon taking the flat-space limit ($\lambda\to 0$), the outer horizon of BTZ is mapped to infinity;
while, the inner horizon takes the role of a cosmological horizon at a finite radius $r_0$. 
In \cite{Hosseini:2015uba}, it was shown that the EE for an interval $(\Delta x,\Delta t)$
in the $\rm{GCFT}_2$ dual to the FSC is given by 
\begin{align}\label{EE FSC}
 S^{\text{FSC}}_{\text{EE}}&=\frac{c_{LL}}{6} \log \left[\frac{\beta _+ }{\pi  \delta }\sinh \left(\frac{\pi \Delta x}{\beta _+}\right)\right]
-\frac{c_{LM}}{6}\frac{\beta _0}{\beta _+}\nn\\&
+\frac{ c_{LM} \pi}{6 \beta _+}\left(\Delta t+\frac{\beta _0 }{\beta _+}\Delta x\right)\coth \left(\frac{\pi  \Delta x}{\beta _+}\right)\ ,
\end{align}
where 
\be
\beta_{+}=\frac{2 \pi }{\hat{r}_+}\, , \qquad
\beta_{0}=\frac{2 \pi  r_0}{\hat{r}_+^2}\, .\label{temps}
\ee
It is worth considering the above expression in different physically relevant regimes,
a summary of these limits can be found in Table \ref{table:Limits} (for more details see \cite{Hosseini:2015uba}).
As we shall see in the following section, the EE for the FSC \eqref{EE FSC} has some interesting
structural similarities with the one corresponding to ALif black holes. 

\begin{sidewaystable}
\centering
\begin{tabular}[t]{@{}l l l@{}}\toprule
Limit & Entanglement entropy & Interpretation \\ \hline
\\[-0.2cm]
$\displaystyle\mu\rightarrow \infty$ &
$\displaystyle \frac{c_{LM}}{6\beta _+}\left[\pi \left(\Delta t+\frac{\beta _0 }{\beta _+}\Delta x\right)
\coth \left(\frac{\pi  \Delta x}{\beta _+}\right)-\beta _0\right]$ & Einstein gravity \\[0.75cm]
$\displaystyle j\rightarrow 0$ &
$\begin{aligned}
& \frac{c_{LL}}{6} \log \left[\frac{\beta _+ }{\pi  \delta }\sinh \left(\frac{\pi \Delta x}{\beta _+}\right)\right]
+\frac{ c_{LM} \pi}{6 \beta _+}\Delta t\coth \left(\frac{\pi  \Delta x}{\beta _+}\right)
\end{aligned}$ & Finite temperature GCFT \cite{Bagchi:2014iea}\\[0.75cm]
$\Delta x\rightarrow \infty$ & $\displaystyle\frac{2\pi r_{0}}{4G}+\frac{2\pi \hat r_{+}}{4 G \mu}$ & Thermal entropy\\[0.75cm]
$\begin{aligned}
& G\rightarrow \infty\, ,\\ & \mu G~\text{fixed}
\end{aligned}$ &
$\displaystyle\frac{c_{LL}}{6}\log \left[\frac{\beta_{+}}{\pi \delta }\sinh \left(\frac{\pi  \Delta x}{\beta_{+} }\right)\right]$ &
Chiral CFT ($\chi$CFT)/ Flat space $\chi$gravity\\[0.75cm]
$\begin{aligned}
& j\rightarrow 2M\, ,\\ & \left(\beta_0=\beta_+=\beta\right)
\end{aligned}$ &
$\begin{aligned}
& \frac{c_{LL}}{6} \log \left[\frac{\beta }{\pi  \delta }
\sinh \left(\frac{\pi \Delta x}{\beta }\right)\right] \\ &
-\frac{c_{LM}}{6}\left[1-\frac{\pi}{\beta }\left(\Delta t+\Delta x\right)\coth \left(\frac{\pi  \Delta x}{\beta }\right)\right]
\end{aligned}$ & Isothermal GCFT (iGCFT)\\[1cm]
\hline
\end{tabular}\caption{Interesting limits of the FSC EE}\label{table:Limits}
\end{sidewaystable}

\section{Entanglement for ALif spacetimes}
\label{EE for ALif}
\subsection{Asymptotically Lifshitz solutions in NMG}

In the following, we consider spacetimes originating from the NMG action \eqref{massiveAc}.
As it turns out, this theory supports a number of interesting 
solutions; in the following we shall study those whose line elements are asymptotically Lifshitz (ALif). For starters, 
Lifshitz spacetimes with an arbitrary dynamical exponent $\nu$
\be\label{Lif. vacuum}
ds^2=-\frac{r^{2\nu}}{L^{2\nu}}dt^2
+\frac{L^2}{r^2} dr^2
+r^2 d\phi^2\, ,
\ee
can be found as long as $\nu$ relates to the NMG parameters as follows
\be\label{parameters}
m^2 L^ 2=\frac{1}{2}\left(\nu^2-3\nu+1\right)\, ,\qquad \lambda L^ 2=-\frac{1}{2}\left(\nu^2+\nu+1\right)\, .
\ee
Hereafter, the reader should keep in mind that the graviton mass $m$ and the dynamical exponent $\nu$ are not independent of each other.
Besides these Lifshitz vacua, for $\nu=1,3$ it is possible to find ALif black hole solutions given by
\be\label{blackholes}
ds^2=-\frac{r^{2\nu}}{L^{2\nu}}\left(1-\frac{8 M G L^2}{r^2}\right)dt^2
+\frac{L^2}{r^2}\left(1-\frac{8 M G L^2}{r^2}\right)^{-1}dr^2\,
+r^2 d\phi^2\, .
\ee
Clearly, for the $\nu=1$ case, the metric is AAdS and Eq.\,\eqref{blackholes} describes a (NMG) BTZ black hole.
On the other hand, for $\nu=3$,
Eq.\,\eqref{blackholes} corresponds to a genuinely ALif black hole \cite{AyonBeato:2009nh}.
Notice that, in view of Eq.\,\eqref{parameters}, these black holes 
appear at the chiral (critical) NMG points $m^2 L^ 2=-1/2$ ($m^2 L^ 2=1/2$) correspondingly.
To compute the holographic EE we will need the Ricci scalars for the Lifshitz vacuum and black hole, which are given by 
\be
R=-\frac{2 \left(\nu ^2+\nu +1\right)}{L^2}\, ,
\ee
and 
\be
R=\frac{2 (\nu -1)^2 r_+^2}{L^2 r^2}-\frac{2 \left(\nu ^2+\nu +1\right)}{L^2}\, ,
\ee
respectively.

\medskip
The ALif black hole described by Eq.\,\eqref{blackholes} has a curvature singularity at $r=0$
and a single event horizon located at $r_{+}=L\sqrt{8 G M}$.
The Hawking temperatures associated to its horizon is given by \cite{AyonBeato:2009nh}
\be\label{temperature}
T=\frac{r_+^3}{2\pi L^{4}}\, ,
\ee
while its Wald entropy reads
\be\label{wald}
S=-\frac{2 \pi r_+}{G}\, .
\ee
The reader might be disconcerted by the fact that the above expression is negative. However, this is a consequence of the fact that in NMG the Einstein-Hilbert term comes with a negative sign. Henceforth, we will be interested in the $\nu=3$ black hole but keep $\nu$ as a bookkeeping device, this will prove useful in finding a general pattern for the Lifshitz EE. 

\subsection{Lifshitz entanglement at zero temperature}

Let us start by computing the holographic EE for generic Lifshitz vacua,  Eq.\,\eqref{Lif. vacuum}. 
We expect this quantity to correspond to the EE for a 1+1 dimensional theory with anisotropic scaling symmetry
\be
t\to\lambda^{\nu} t\;,\qquad x\to\lambda x\,,
\ee
we refer to $\nu$ as the \emph{dynamical exponent}.
To compute the holographic EE, we use the functional \eqref{EE H1}.
Let us suppose that, at a constant time slice, the entangling curve into the bulk has the form
$(r,\phi(r))$. Using this as well as the metric \eqref{Lif. vacuum},
the induced metric $h$ and $R_{||}$ are given by
\begin{align}\label{induced vacuum}
h=r^2 \phi'(r)^2+\frac{L^2}{r^2}\, ,\qquad
R_{||}=\frac{(\nu -1) \nu }{L^2+r^4 \phi '(r)^2}-\frac{2 \nu ^2+\nu +1}{L^2}\, ,
\end{align}
while the extrinsic curvature contraction reads
\begin{align}
{\cal K}^{2}(r)&
=-\frac{\left[L^2 r^3 \phi ''(r)+3 L^2 r^2 \phi '(r)+r^6 \phi '(r)^3\right]^2}{L^2 \left[L^2+r^4 \phi '(r)^2\right]^3}\, .
\end{align}
We would like to point out an important difference between the RT and the higher-curvature holographic EE prescriptions. 
If we employ the RT prescription for a fixed time slice and we consider two line elements that differ only in their $g_{tt}$ components, then we notice that 
the computation is insensitive to that difference. However, this is not the case in the presence of higher derivative corrections, as can be seen from $R_{||}$ in Eq.\,\eqref{induced vacuum} which has a nontrivial dependence on $\nu$ which appears only in the time component of the metric \eqref{Lif. vacuum}.

\medskip
Plugging these into the functional \eqref{EE H1} and carrying out the variation we find a complicated 
fourth-order equation of motion for $\phi(r)$. Fortunately, the solution of this equation satisfying the
free-kick condition \cite{Hosseini:2015uba} matches exactly  the profile traced by a geodesic curve. 
To find the geodesics in the Lifshitz background one simply needs to extremize 
the functional
\be\label{length}
I(A)=
\int_{\Sigma_A} d^{d-1}y\sqrt{h}\, ,
\ee
with the $h$ given in \eqref{induced vacuum}.
Setting $L=1$, we find that the profile of a geodesic, anchored to the boundary, reaching down to a radius $r_*$ is simply
\be\label{geo}
\phi(r)=\frac{\sqrt{r^2-r_*^2}}{r r_*}\, .
\ee
From the boundary point of view one would wish to specify the size of the entangling region $\tilde\phi$ as a free parameter instead of the bulk depth $r_*$.
It is clear that these two quantities are related by  
\be\label{boundary bulk}
\tilde\phi_{\rm geodesic}(r_*)= \phi(r,r_*)\Big|_{r\to\infty}\, .
\ee
To lighten the notation, we drop the subscript \lq\lq{}geodesic\rq\rq{} when no confusion arises.

\medskip
Notice that although the relevant extremal curves are the same as in Einstein gravity it is not
their length that we must consider, instead we ought to insert \eqref{geo} into 
\eqref{EE H1} and compute the integral. Thus, we find that the holographic EE corresponding to 
the Lifshitz vacuum in NMG is given by
\begin{align}
S_{\rm EE}&=\frac{2}{4 G}
\int_{r_{*}}^{r_{\delta }}
\frac{r^2 \left(2 m^2+\nu(\nu -1)+1\right)-2 \nu (\nu -1) r_*^2}{2 m^2 r^2 \sqrt{r^2-r_*^2}}\, dr\nonumber\\
&=\frac{1}{4 G m^2}\left[\left(2 m^2+\nu(\nu -1)+1\right) \log \left(\sqrt{r^2-r_*^2}+r\right)
-\frac{2 \nu (\nu -1) \sqrt{r^2-r_*^2}}{r}\right]
\Bigg|_{r_{*}}^{r_{\delta }}\, ,
\end{align}
where we introduced an ultraviolet cutoff $r_\delta\gg 1$.
Replacing $r_*$ and $r_\delta$ using  
\be
r_{*}=\frac{1}{\tilde\phi}\, ,\qquad r_{\delta}=\frac{1}{\delta}\, ,
\ee
we find that the EE for the Lifshitz vacua is
\be\label{Lif Vac EE}
S_{\rm EE}=\frac{1}{2 G}\left[\left(1+\frac{1}{2m^2}\right)+\frac{\nu(\nu-1)}{2m^2}\right]
\log \left(\frac{2 \tilde \phi }{\delta }\right)
-\frac{\nu(\nu-1)}{ 2 G m^2}\, .
\ee
Finally, using Eq.\,\eqref{parameters} we can write the above in terms of the dynamical exponent
\be\label{Lif Vac EE1}
S_{\rm EE}=\frac{(\nu-1)^2}{ G \left[\nu(\nu-3)+1\right]}\left[
\log \left(\frac{ \Delta x }{\delta }\right)
-\frac{\nu}{\nu-1}\right]\, ,
\ee
where $\Delta x=2 \tilde \phi$ is the size of the entangling region. 

\subsection{Lifshitz entanglement at finite temperature}

In the present section we will make a proposal for the general form of the EE for (1+1)-dimensional Lifshitz field theories at finite temperature. From a holographic standpoint, this amounts to performing a computation analogous to that in the previous section, now, with a black hole in the bulk. 
Hence, we consider a metric of the form \eqref{blackholes}; for such a spacetime 
the quantities entering the EE functional \eqref{EE H1} are displayed below. 
The induced metric is given by
\be
h=\frac{L^2}{r^2-r_{+}^2}+r^2 \phi '(r)^2\, ,
\ee
while the contribution from the ambient Riemann curvature is
\begin{align}
R_{||}=&\left[L^4 r^2+L^2 r^4 \left(r^2-r_+^2\right) \phi '(r)^2\right]^{-1} \Big\{L^2 \left[\nu(\nu -1) r_+^2-(\nu +1)^2 r^2\right]\nn\\  
&-r^2 \left(r^2-r_+^2\right) \left[\left(2 \nu ^2+\nu +1\right) r^2+(\nu(5-2 \nu ) -3) r_+^2\right]
\phi '(r)^2\Big\}\,,
\end{align}
and the trace of the extrinsic curvature reads
\begin{align}
{\cal K}^{2}(r)&
=-\frac{\left[L^2 r \left(r^2-r_+^2\right) \phi ''(r)+L^2 \left(3 r^2-2 r_+^2\right) \phi '(r)
+\left(r^3-r r_+^2\right)^2 \phi '(r)^3\right]^2}{L^2 \left[L^2+\left(r^4-r^2 r_+^2\right) \phi '(r)^2\right]^3}\, .
\end{align}
Once again, 
the problem of extremizing the functional \eqref{EE H1} with free-kick conditions boils down to finding the appropriate geodesics anchored to the boundary, which for the black hole read
\be\label{geobtzphi}
\phi(r)= \frac{1}{r_+}\,\text{arccosh}\left(\sqrt{\frac{r_*^2-\left(\frac{r_+r_*}{r}\right)^2}{r_*^2-r_+^2}}\right)\, .
\ee
The next step is to plug this profile back into  Eq.\,\eqref{EE H1} and compute the integral
\begin{align}
S_{\rm EE}=\frac{2}{4 G}
\int_{r_{*}}^{r_{\delta }}
\frac{r^4 \left(2 m^2+\nu (\nu -1) +1\right)
-(\nu -1) \left[(\nu -3) \left(r^2-2 r_*^2\right) r_+^2+2 \nu  r^2 r_*^2\right]}
{2 m^2 r^3 \sqrt{\left(r^2-r_+^2\right) \left(r^2-r_*^2\right)}}\, dr\, ,
\end{align}
which fortunately can be written in terms of hypergeometric functions as
\begin{align}
S_{\rm EE}&=\frac{1}{8 G m^2 r^2}\Bigg[
2 r^2 \left(2 m^2+\nu (\nu -1) +1\right) \log \left(\sqrt{r^2-r_+^2}+\sqrt{r^2-r_*^2}\right)\\&
+(\nu -1) (\nu +3) r_*^2 F_1\left(1;\frac{1}{2},\frac{1}{2};2;\frac{r_+^2}{r^2},\frac{r_*^2}{r^2}\right)
-2 (\nu -3) (\nu -1) \sqrt{\left(r^2-r_+^2\right) \left(r^2-r_*^2\right)}\Bigg]\nn
\Bigg|_{r_{*}}^{r_{\delta }}\, ,
\end{align}
where $r_\delta\gg 1$ is an ultraviolet cutoff. Finally, if
we replace $r_*$ and $r_\delta$ using 
\be
r_{*}=r_+ \coth \left(r_+ \tilde{\varphi }\right)\, ,\qquad r_{\delta}=\frac{1}{\delta}\, ,
\ee
we find
\begin{align}\label{EE lif temp}
 S_{\rm EE} =& \frac{1}{2 G}\left[\left(1+\frac{1}{2m^2}\right)+\frac{\nu(\nu-1)}{2m^2}\right]
\log \left[\frac{2}{\delta  r_+}\sinh \left(r_+ \tilde\phi\right)\right]\nonumber\\
   &-\frac{\nu-1}{4 G m^2} \big[(\nu+3) (r_+ \tilde\phi) \coth \left(r_+ \tilde\phi\right)+(\nu-3)\big] \, . 
\end{align}
The thermal entropy of the system can be recovered
once the entangling region becomes large.
Applying this limit to \eqref{EE lif temp}, we obtain
\be
S=\frac{\pi r_+}{ 2 G}\frac{(\nu -1)(\nu -5)}{\nu(\nu -3)+1}\, ,
\ee
in agreement with \eqref{wald}, for $\nu=3$.

\medskip
Recall that the the metric \eqref{blackholes} is a bona-fide solution of the NMG eom only for $\nu=3$.
Nevertheless, we claim that \eqref{EE lif temp} captures the finite temperature EE
for field theories with arbitrary dynamical exponent $\nu$; we shall provide an argument in favor of this idea shortly. 

\medskip
It is instructive to express the EE \eqref{EE lif temp} in terms of parameters corresponding to the putative field theory dual.
First, we introduce the quantities
\begin{align}
&c_{\rm LL}(\nu)=2 c^{\rm AdS}+\frac{3\nu(\nu-1)}{2 G m^2}\, ,
\quad c_{\rm LX}(\nu)=\frac{\nu}{1-\nu}\,c_{\rm LL}(\nu)\, ,
\end{align}
where 
\be
c^{\rm AdS}=\frac{3}{2G}\left(1+\frac{1}{2m^2}\right)\, ,
\ee
is the central charge corresponding to the AdS vacuum in NMG. 
Recall that $m$ is not an independent parameter but is related to $\nu$ via equation \eqref{parameters}. In terms of $\nu$ the above quantities are given by 
\begin{align}
&c_{\rm LL}(\nu)= \frac{6}{G}\left[\frac{(\nu-1)^2}{\nu(\nu-3)+1} \right]\, ,
\quad c^{\rm AdS}=\frac{3}{2G}\left[\frac{(\nu-1)(\nu-2)}{\nu(\nu-3)+1}\right]\, .
\end{align}
Notice that in the relativistic limit ($\nu\rightarrow 1$) we find $c^{\rm AdS}=0$, corresponding to the NMG chiral point;
which is believed to be dual to a logarithmic CFT (see \cite{Grumiller:2009sn} and references therein). 

\medskip
Finally, using Eq.\,\eqref{temperature} we write the result \eqref{EE lif temp} in terms of the inverse Hawking temperature 
\be
\beta=\frac{2\pi}{r_+^\nu}\,,
\ee
and find 
\begin{align}\label{finite temp}
S_{\rm EE}=&\frac{c_{\rm LL}(\nu)}{6}\log \left\{\frac{2}{\delta}\left(\frac{\beta}{2\pi}\right)^{1/\nu}
\sinh \left[\frac{\Delta x}{2}\left(\frac{2\pi}{\beta}\right)^{1/\nu}\right]\right\}\\
  &+\frac{c_{\rm LX}(\nu)}{12}
\left\{\frac{(\nu+3)}{\nu} \frac{\Delta x}{2}\left(\frac{2\pi}{\beta}\right)^{1/\nu}
\coth \left[\frac{\Delta x}{2}\left(\frac{2\pi}{\beta}\right)^{1/\nu}\right]\nn
+\frac{\nu-3}{\nu}\right\}\, ,
\end{align}
where $\Delta x=2\tilde\phi$ corresponds to the length of the entangling interval. Observe that in the zero temperature limit the above expression takes (after some non-trivial cancellations) the form 
\be\label{zero t}
S_{\rm EE}=\frac{c_{\rm LL}(\nu)}{6}\log \left(\frac{\Delta x}{\delta }\right)
+\frac{c_{\rm LX}(\nu)}{6}\,  ,
\ee
in agreement with Eq.\,\eqref{Lif Vac EE1}. Let us make some important clarifications. 
The zero temperature result Eq.\,\eqref{Lif Vac EE1} was obtained, via holography, for arbitrary dynamical exponents. On the other hand, in this work we can invoke holography for the finite temperature EE \eqref{finite temp} only for $\nu=3$. However, the fact that the RHS of  Eq.\,\eqref{finite temp} reduces exactly to \eqref{Lif Vac EE1} in the zero temperature limit leads us to conjecture that 
\eqref{finite temp} is valid also for systems with $\nu\neq3$. We plan to address this question in future work.

\medskip
To close this section, we would like to point out a remarkable structural similarity between the Lifshitz EE \eqref{finite temp}
and the EE for a GCFT dual to the isothermal limit of the FSC (see Table \ref{table:Limits}).
Notice that both of the underlying symmetry algebras [\eqref{LifA}, \eqref{GCA}] corresponding to these theories contain a
Virasoro subalgebra, which accounts for the logarithmic contributions to the EE.
We expect this logarithmic contribution to be a general feature of the EE for theories with a Virasoro symmetry subalgebra. The reason being that the EE for an interval can be calculated via a two-point function of primary fields, called twist fields, inserted at the endpoints of the entangling region. The Ward identities corresponding to the Virasoro subalgebra impose strong constraints on the form of these two-point functions, which lead to a logarithmic contribution in the EE.
One might wonder whether the $\coth$ contributions are a signature of the breakdown of Lorentz invariance
in nonrelativistic CFT's. Moreover, the constants $c_{LL}$ and $c_{LM}$ appearing in the FSC EE
correspond to the central extensions appearing in the algebra \eqref{GCA}.
Hence, it is tempting to regard the quantities $c_{LL}(\nu)$ and $c_{LX}(\nu)$
as manifestations of the anomaly coefficient in a $\rm{Lif}_2$ field theory. 

\section{Mutual information in nonrelativistic field theories}
\label{MI}
In this section, we study another interesting measure of entanglement called \textit{mutual information}.  Given two disjoint subsystems $A$ and $B$, their mutual information is defined 
by
\be 
I(A:B)=S_A+S_B-S_{A\cup B}\,,
\ee
where $S_A$, $S_B$ and $S_{A\cup B}$ are the EE's for $A$, $B$ and $A\cup B$ respectively.
There are a number of nice properties of $I(A:B)$, for example, it is a finite positive semi-definite quantity.
The mutual information captures the total amount of correlations between $A$ and $B$ \cite{QI}. Moreover,
it provides an upper bound for the connected two-point functions between operators supported on $A$ and $B$.

\begin{figure}[t]
    \centering
    \includegraphics[scale=0.45]{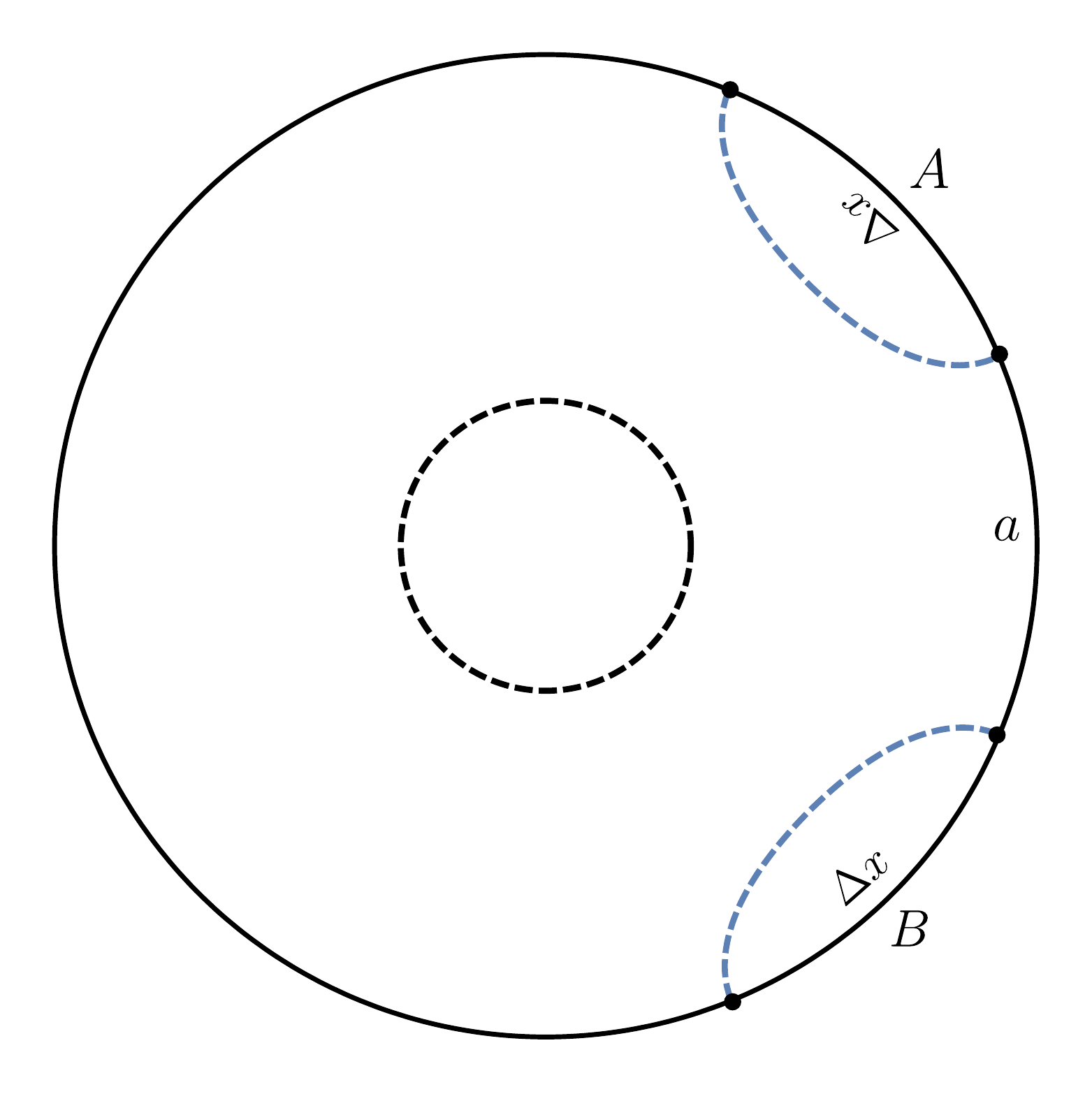}\quad
    \includegraphics[scale=0.45]{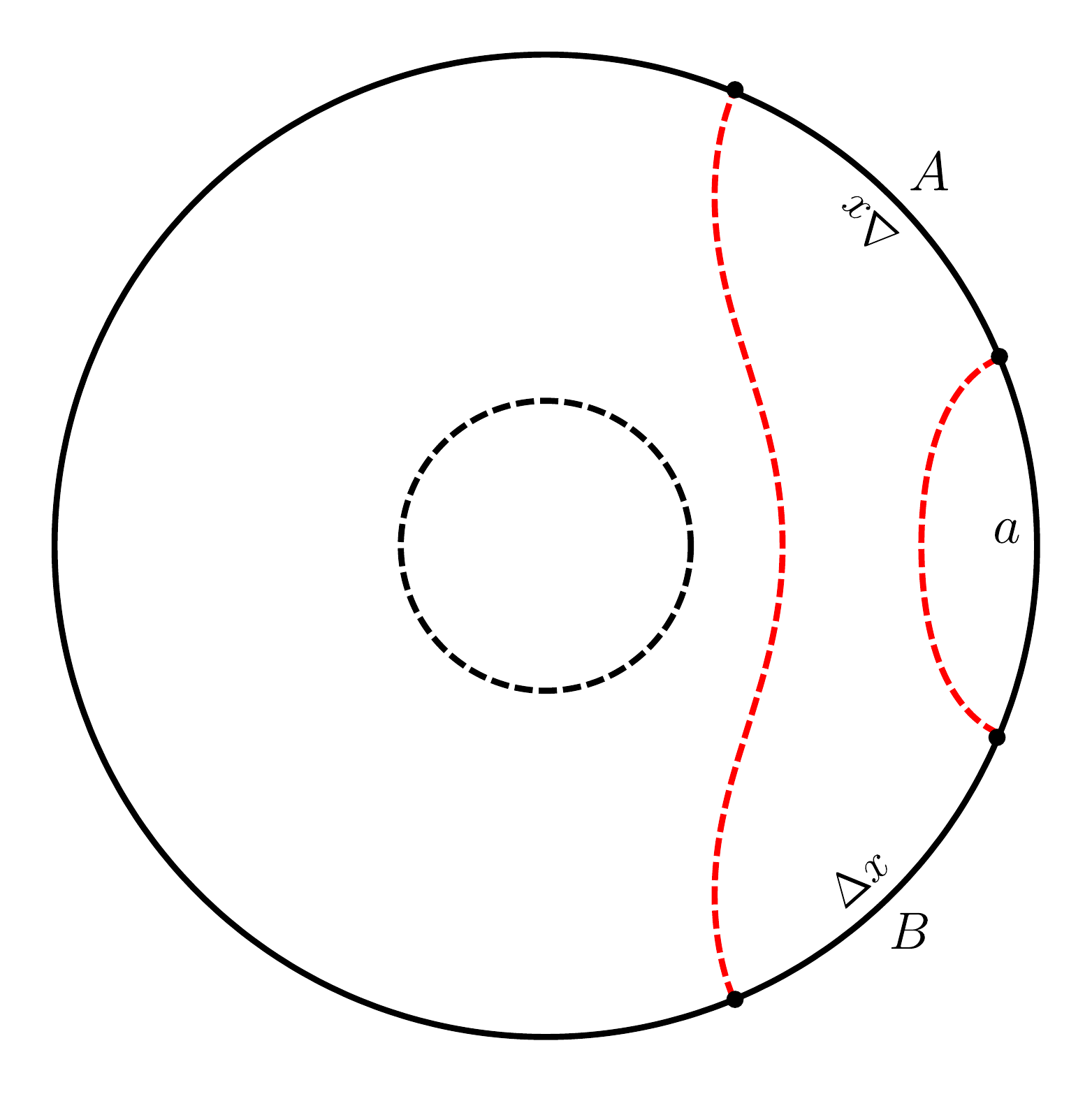}
\caption{Possible minimal surfaces encoding the entanglement entropy for $S_{A\cup B}$, where we used $r \to \arctan (r)$
to compactify the radial coordinate.}\label{MI1}
\end{figure}

\medskip
In the following, let $A$ and $B$ be intervals of length $\Delta x$ separated by a distance $a>0$ as depicted in Fig.\,\ref{MI1}. 
For a single interval, the finite temperature EE's in CFT's, GCFT's and $\rm{Lif}_2$ theories are known, so the $S_A$ and $S_B$ contributions
to the mutual information can be readily obtained. Hence, the non-trivial step to find the mutual information is the computation of the EE for the disjoint subsystem $S_{A\cup B}$. The extremal surface anchored 
to $A\cup B$ can take either of the two shapes\footnote{In global coordinates there are four possible phases instead of two, see \cite{Ben-Ami:2014gsa} for more details.} displayed in Fig.\,\ref{MI1}. To solve this problem correctly,
one must choose the $\Sigma$ giving the minimal value once substituted back into the functional \eqref{EE H1}.
Interestingly, this implies a first order phase transition in the mutual information \cite{Headrick:2010zt,Fischler:2012ca,Fischler:2012uv}.
Indeed, when the configuration on the right of Fig.\,\ref{MI1} is the relevant one we obtain
\be
I(A:B)=2S(\Delta x)-S(a)-S(2\Delta x+a)\,,
\ee
on the other hand, if we are compelled to choose the figure on the left we have
\be
I(A:B)=0\,.
\ee

\begin{figure}[t]
    \centering
    \includegraphics[scale=0.64]{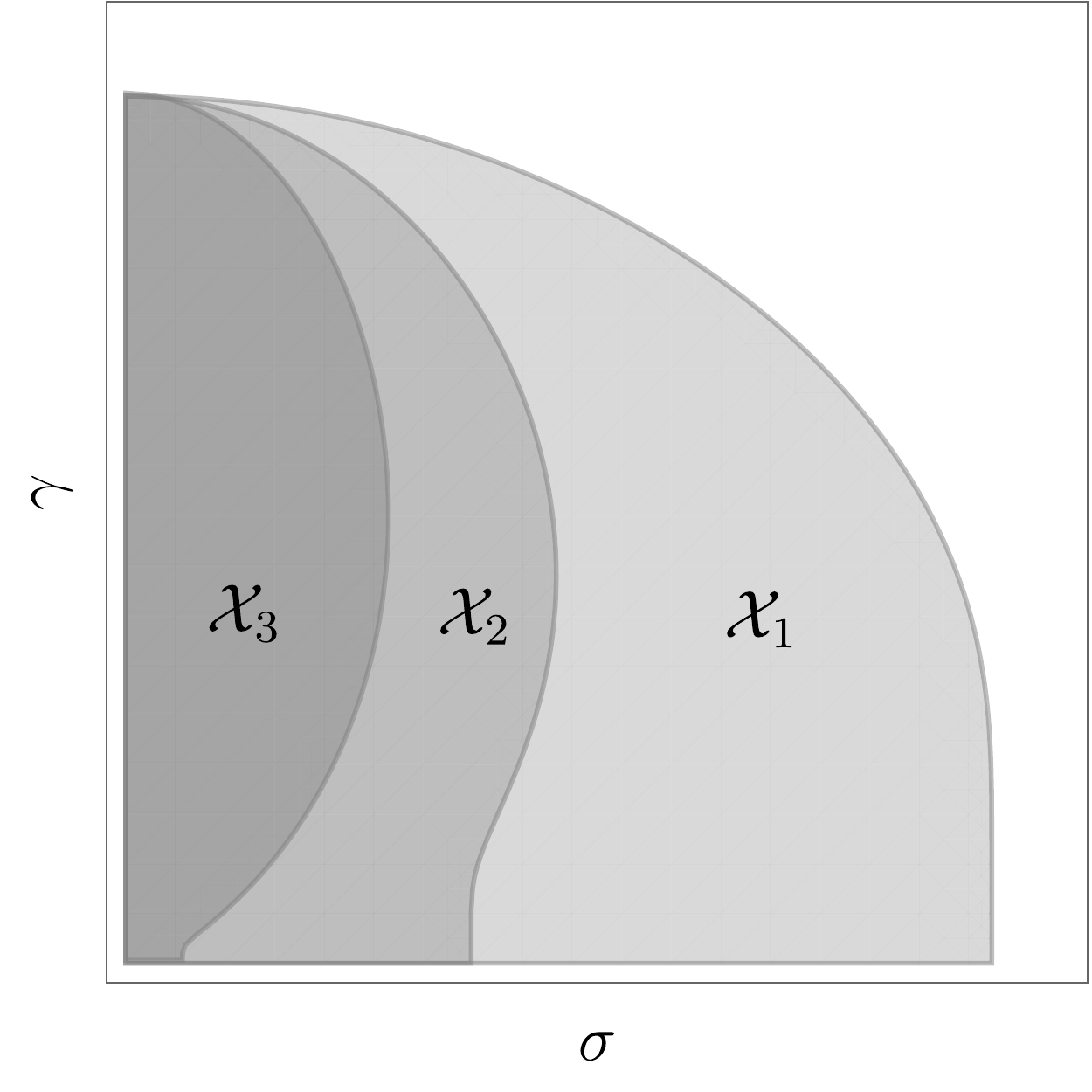}
\caption{Mutual information phase diagram for $\rm{GCFT}_2$, where $(\sigma, \gamma)$ are defined by Eq.\,\eqref{dimensionless}, and ${\cal X}=c_{LM}/c_{LL}$. In the above figure, the gray areas have non-vanishing mutual information and we consider different values of the ratio ${\cal X}$,  $0={\cal X}_1<{\cal X}_2<{\cal X}_3$.  }\label{MI-phase}
\end{figure}

\medskip
Now, consider the mutual information for an 
isothermal GCFT ($\rm{iGCFT}$, see Table\,\ref{table:Limits}).
For such a theory, the mutual information in the non-vanishing phase reads
\begin{align}
I(A:B)&=\frac{c_{LL}}{6}\log \left\{\frac{
\sinh ^2\vast(\frac{\pi \Delta x}{\beta }\vast)}
{\sinh \vast(\frac{\pi a}{\beta }\vast)
\sinh \left[\frac{\pi  (2 \Delta x+a)}{\beta }\right]
}\right\}\\&
+\frac{\pi c_{LM}}{6\beta}
\left\{2 \Delta x\coth \left(\frac{\pi \Delta x}{\beta }\right)
-(2 \Delta x+a) \coth \left[\frac{\pi  (2 \Delta x+a)}{\beta }\right]
-a \coth \left(\frac{\pi a}{\beta }\right)\right\}\, .\nn
\end{align}
For upcoming discussions, it is convenient to introduce the dimensionless quantities
\be\label{dimensionless}
\sigma=\frac{a}{\beta}\, \qquad\gamma=\frac{a}{\Delta x}\, .
\ee
The disentangling phase transition in the $(\sigma, \gamma)$-plane occurs when crossing the curve determined by the equation
\begin{align}
&-\log \left\{\frac{
\sinh ^2\vast(\frac{\pi \sigma}{\gamma }\vast)}
{\sinh \left(\pi\sigma\right)
\sinh \left[\pi\left(\frac{2\sigma}{\gamma} +\sigma \right)\right]
}\right\}=\nn\\&
\pi {\cal X}
\left\{ \frac{2\sigma}{\gamma}\coth \left(\frac{\pi \sigma}{\gamma }\right)
-\left(\frac{2\sigma}{\gamma} +\sigma\right) \coth \left[\pi\left(\frac{2\sigma}{\gamma} +\sigma \right)\right]
-\sigma\coth \left(\pi\sigma\right)\right\}\, ,
\end{align}
where ${\cal X}=c_{LM}/c_{LL}$.
The resulting phase diagram is depicted in Fig.\,\ref{MI-phase}, where the gray areas represent the regions with non-vanishing 
mutual information. 

\medskip
Let us make some important comments regarding Fig.\,\ref{MI-phase}. First, observe that the gray region for ${\cal X}=0$ is identical to that corresponding to 
a BTZ black hole as discussed in \cite{Fischler:2012uv}. Moreover, notice that at zero temperature, corresponding to $\sigma=0$, all the transition boundaries converge to a universal 
value $\gamma=\tilde\gamma=\sqrt{2}-1$. This implies that, both for $\rm{CFT}$'s and $\rm{iGCFT}$'s, there is a unique threshold for the ratio size to separation in Fig.\,\ref{MI1}, 
after which the mutual information vanishes. On the other hand, as we heat up the system there are qualitative changes in the phase diagram depending on the value of ${\cal X}$. Instead, if we wish to consider a $\rm{Lif}_2$ field theory some differences must be taken into account. First of all, we should exchange $\sigma$ for 
\be
\sigma(\nu)=\frac{a}{\beta^{1/\nu}}\,.
\ee
Once more, the behavior of the phase diagram is controlled by the ratio ${\cal X}=c_{LX}/c_{LL}$, which now is a function of $\nu$. Whenever ${\cal X}$ is non-negative, the aspect of the phase diagram is totally analogous to that of an $\rm{iGCFT}$. Otherwise, if ${\cal X}$ takes negative values there is a point at which the non-vanishing mutual information region becomes unbounded, we believe this to be a consequence of non-unitarity.

\section{Conclusions and outlook}
\label{Conclusions and outlook}

Systems with anisotropic scaling properties appear frequently
in quantum and statistical field theory of condensed matter systems. In this work we have studied certain entanglement 
properties for such theories. We have calculated the EE for a class of Lif$_2$ field theories both at zero \eqref{zero t} and finite temperature \eqref{finite temp}. 
As far as we know, this is the first instance of such computation. Moreover, it was noticed that 
these expressions display an interesting structural similarity 
with the EE corresponding to iGCFTs. This observation leads us to conjecture the form of the Lif$_2$ anomaly coefficient dual to the Lifshitz vacuum in NMG.
Finally, we investigated the phase space of mutual information for the aforementioned theories. For iGCFTs, the curves defining the boundary between the two mutual information phases display a compelling behavior as one tunes the ratio of the central charges, a similar behavior is observed for Lif$_2$ field theories.  
  
\medskip
Let us finish by pointing out some potentially interesting directions. We believe that these results might be understood solely in terms of Lifshitz (or Schr\"odinger) Ward identities on the plane and cylinder. It would be interesting to compute the EE and mutual information for simple field theories with Lifshitz scaling. Another puzzle that might be interesting to address is the holographic renormalization for Lifshitz backgrounds in NMG. This could shed light on the nature of the Lif$_2$ anomaly coefficient. Furthermore, we would like to explore the behaviour of EE under the renormalization group flow connecting nonrelativistic and relativistic fixed points. Moreover, it would be interesting to investigate the analogue of this questions in higher dimensions.

\section*{Acknowledgments}
We are grateful to Reza Fareghbal, Kevin Goldstein and Vishnu Jejjala for commenting on a draft of this article.
We would like to thank Pawel Caputa, Niels Obers, Sergey Solodukhin, Alessandro Tomasiello and Alberto Zaffaroni
for illuminating conversations and correspondence.
Additionally, we wish to acknowledge Silvia Ferrario Ravasio for her kindness answering our Latex questions.
The research of AVO is supported by the University Research Council of the University of the Witwatersrand.
The work of SMH is supported in part by INFN. SMH and AVO wish to thank the hospitality of the ICTP
in Trieste where part of this research was conducted. 

\begin{appendix}

\section{A brief review of massive gravity in 2+1 dimensions}
\label{AppA}

The action of (2+1)-dimensional massive gravity can be written as \cite{Deser:1982vy,Bergshoeff:2009hq}
\be\label{massiveAc}
S=\frac{1}{16\pi G}\int d^{3}x \sqrt{-g}\left[R-2\lambda
+\frac{1}{m^{2}}K+\frac{1}{2\mu}{\rm CS}(\Gamma)\right],
\ee
where
\begin{align}
K=R_{\mu\nu}R^{\mu\nu}-\frac{3}{8}R^{2}\, ,\qquad
{\rm CS}(\Gamma)=\varepsilon ^{\alpha \beta \gamma}\Gamma _{\;\alpha \sigma }^{\rho }
\left(\partial _{\beta }\Gamma _{\;\gamma \rho}^{\sigma }
+\frac{2}{3}\Gamma _{\;\beta \eta }^{\sigma }\Gamma _{\;\gamma \rho}^{\eta }\right)\, .
\end{align}
Here, $\lambda$ is a cosmological parameter and $m,\,\mu$ are mass parameters.
The variation of \eqref{massiveAc} reads
\be\label{fieldeq}
R_{\mu \nu }-\frac{1}{2}Rg_{\mu \nu }
+\lambda g_{\mu \nu }+\frac{1}{2m^{2}}K_{\mu \nu }
+\frac{1}{\mu}C_{\mu \nu }=0\, ,
\ee
where $C_{\mu \nu}$ is the Cotton tensor given by
\be
C_{\mu \nu}=\varepsilon _{\mu }^{~\alpha \beta }\nabla_{\alpha }
\left(R_{\beta \nu }-\frac{1}{4}g_{\beta\nu}R\right)\, ,
\ee
while $K_{{\mu \nu }}$ reads
\begin{equation}
K_{\mu \nu }=2\square {R}_{\mu \nu }-\frac{1}{2}\left(\nabla _{\mu }\nabla _{\nu }{R}+g_{\mu \nu}\square {R}\right)
-8R_{\mu \alpha}R^{\alpha}_{~\nu}
+\frac{9}{2}RR_{\mu \nu }
+g_{\mu\nu }\left(3R_{\alpha \beta }R^{\alpha \beta}-\frac{13}{8}R^{2}\right)\, ,
\end{equation}
and fulfills $K=g^{\mu \nu }K_{\mu\nu }$.
In this work we wish to study the EE for (2+1)-dimensional ALif black holes.
Such black holes have been found in \cite{AyonBeato:2009nh} in the context of NMG,
namely the theory defined by sending $\mu\to\infty$ in action \eqref{massiveAc} \cite{Bergshoeff:2009hq}.
This theory is diffeomorphism and parity invariant.

\label{galilean}\section{Galilean conformal algebra}
\label{AppB}
The centrally extended Galilean conformal algebra in 1+1 dimensions,
which is the case for the present work, can be written as
\begin{align}\label{GCA}
  \nonumber &[L_m,L_n]=(m-n)L_{n+m}+{c_{LL}\over 12}m(m^2-1)\delta_{m+n,0}\,,\\
  \nonumber &[L_m,M_n]=(m-n)M_{n+m}+{c_{LM}\over 12}m(m^2-1)\delta_{m+n,0}\, ,\\ 
            &[M_m,M_n]=0\, ,
  \end{align}
for $n\,,m\in\mathbb{Z}$.
The GCA$_2$ is generated by the set of conformal isometries of Galilean spacetimes.
The vector fields that span this algebra are given \cite{Bagchi:2009my,Bagchi:2012cy}
\begin{align}\label{GCAgen}
L_n=-(n+1)x^nt\pd_t-x^{n+1}\pd_x\, ,\qquad M_n=x^{n+1}\pd_t\, .
\end{align}
The class of field theories endowed with the symmetry algebra \eqref{GCA} are known as Galilean conformal field theories (GCFT$_2$).
States in a GCFT$_2$ are labeled by the eigenvalues under $L_0$ and $M_0$. The primary states are those annihilated by all generators with $n>0$.
One can build up a families of operators by acting on
primaries with the creation operators $L_{-n}$ and $M_{-n}$ for all $n>0$.

\section{Lifshitz algebra}
\label{AppC}
Defining the generators
\begin{align}
&L_n= -\left[t^{n+1}\pd_t+\frac{x}{\nu}(n+1)t^n\pd_x\right]\, ,\qquad \;\,n\in \mathbb{Z}\nn\, ,\\
&X_a=-t^{a+1/2}\pd_x\, ,\hspace{4cm} a\in \mathbb{Z}+\frac{1}{2}\, ,
\end{align}
we end up with the centrally extended Lifshitz algebra in 1+1 dimensions
\begin{align}\label{LifA}
&\left[L_n,L_m\right]= (n-m)L_{m+n}+\frac{c_{LL}(\nu)}{12}n(n^2-1)\delta_{m+n,0}\nn\, ,\\
&\left[L_n,X_a\right]= \left(\frac{n}{\nu}-a+\frac{2-\nu}{2\nu}\right) X_{n+a}\nn\, ,\\
&[X_a,X_b]=0\, .
\end{align}
This algebra can be obtained by looking at the
asymptotic symmetries for Lifshitz spacetimes \eqref{Lif. vacuum} \cite{Compere:2009qm}.

\end{appendix}

\end{document}